\documentclass[12pt]{article}
\usepackage{mathtools}

\newcommand{\be}{\begin{equation}}
\newcommand{\ee}{\end{equation}}
\newcommand{\bea}{\begin{eqnarray}}
\newcommand{\eea}{\end{eqnarray}}
\newcommand{\betaa}{b}

\begin{document}

\large

\title{ Euler's Reflection Formula, Infinite Product Formulas, and the Correspondence Principle of Quantum Mechanics} 

\begin{center}

\large {\bf Euler's Reflection Formula, Infinite Product Formulas, \\ and the Correspondence Principle of Quantum Mechanics } 

\vskip .3cm

\normalsize

Tamar Friedmann\footnote{Email address: tfriedma@colby.edu} 

\small
{\it Department of Mathematics and Statistics \\ Colby College, Waterville, ME, United States}

\vskip .3cm

\normalsize
Quincy Webb\footnote{Email address: qwebb@smith.edu}
\small

{\it Department of Mathematics and Statistics \\
Smith College, Northampton, MA, United States }
\end{center}

\begin{abstract}

We generalize the derivation of the Wallis formula for $\pi$ from a variational computation of the spectrum of the Hydrogen atom. 
We obtain infinite product formulas  for certain combinations of gamma functions, which include irrational numbers such as $\sqrt 2$ as well as some nested radicals. We also derive Euler's reflection formula for reciprocals of positive even integers. We show that Bohr's correspondence principle allows us to derive our product formulas and the reflection formula without the need for the limit definition of the gamma function. 
\end{abstract}
\vskip .7cm


\normalsize

\subsection*{Introduction}

In \cite{FH1}, the Wallis formula for $\pi$, 
\be \label{Wallis}{\pi \over 2}={2 \cdot 2\over 1\cdot 3}\; {4 \cdot 4\over 3\cdot 5}\; {6 \cdot 6\over 5\cdot 7}\cdots ,\ee 
was derived from a variational computation of the spectrum of the hydrogen atom. 
A trial wave function was used to approximate the energy levels of the hydrogen atom. Then Bohr's correspondence principle  was used to take a classical limit in which the approximate and exact results approach each other, ultimately leading to the Wallis formula.

It was shown in \cite{CS} that the Wallis formula for $\pi$ obtained in \cite{FH1} can be obtained using a  different trial wave function  than the one used there; it was also shown that the sum of some infinite series involving $\pi$ can be obtained in a similar approach. It was then shown in \cite{CG} that the Wallis formula can also be obtained from a variational computation of the spectrum of a different potential that appears in quantum mechanics, namely the harmonic oscillator potential.  

In this paper, we generalize the trial wave function of \cite{FH1} to obtain a systematic set of infinite product formulas for certain combinations of gamma functions. 
While these infinite product formulas can be obtained from previously-known mathematical formulas, here they  are being derived for the first time by using a variational computation and a trial wave function within the framework of quantum mechanics. 
We follow \cite{FH1} closely in presenting the derivation, first in three dimensions and then in $N$ dimensions. It appears that some of the formulas obtained in the present paper can also be obtained from a similar computation involving the harmonic oscillator potential, using the observations made in \cite{CG}. 

Using these product formulas, we also derive Euler's reflection formula, $\Gamma (1-z) \Gamma(z)= {\pi \over \sin \pi z } $, for $z=1/2s$, $s$ a positive integer. Again, this formula has not been derived from physics until now. 

We conclude by discussing the role of different definitions of $\Gamma (z)$ in mathematical derivations of the product formulas and their relation to the role played by Bohr's correspondence principle in our derivation.

\subsection*{Product formulas from the 3-dimensional hydrogen atom}

The Schr{\"o}dinger equation for the hydrogen atom  is given by
$$ H \psi =\left ( -{ \hbar^2\over 2m}\nabla ^2 - {e^2\over r} \right ) \psi =E\psi ,
$$
with the corresponding radial equation obtained by separation of variables being
$$H(r) R(r)=\left [-{ \hbar^2\over 2m}\left ( {d^2\over dr^2}+{2\over r}{d\over dr}-{\ell(\ell +1)\over r^2}\right ) -{e^2\over r} \right ] R(r)=ER(r).
$$

Now we use the following trial wave function:

\be \label{trial} \psi_{\alpha \betaa \ell m} =r^\ell e^{-\alpha r^\betaa} Y^m_\ell(\theta, \phi) , \ee
where $\alpha >0$ and $\betaa >0$ are real parameters and the $Y^m_\ell (\theta, \phi)$ are the usual spherical harmonics. In \cite{FH1}, $\betaa$ was fixed at $\betaa=2$. 
The expectation value of the Hamiltonian is found to be given by
\bea \label{ExpectH}\langle H\rangle _{\alpha  \betaa \ell}&\equiv& {\langle \psi_{\alpha \betaa \ell m}| H(r)  |\psi_{\alpha \betaa \ell m} \rangle \over \langle \psi_{\alpha \betaa \ell m} |\psi_{\alpha \betaa \ell m}\rangle} \nonumber \\
&=& 
\left . {{\hbar ^2\over 2m}\Gamma\left( 2\ell+1+\betaa \over \betaa \right ) \left ( \betaa (2\ell +\betaa +1)\over 4 \right ) 
(2\alpha)^{2 / \betaa} 
-e^2{\Gamma \left (2\ell +2\over \betaa \right ) }(2\alpha)^{1/\betaa}\; \over
\Gamma \left ( 2\ell +3\over \betaa\right )}\right . ,
\eea
where we used the orthonormality of spherical harmonics. Minimizing with respect to $\alpha$ gives

\be \label{min}\langle H \rangle _{min}^{\ell, b} = -{me^4\over \betaa \hbar ^2}{1\over ({2\ell + \betaa +1\over 2})}\; {\left (\Gamma({2\ell +2\over \betaa})\right )^2\over \Gamma ({2\ell +3\over \betaa})\Gamma ({2\ell +\betaa + 1\over \betaa})},\ee
which is an upper bound for the lowest energy state with the given value of $\ell$. 

The exact value for the lowest energy eigenstate for a given $\ell$ is 
\be \label{exact} E_{0,\ell}=-{me^4\over 2\hbar ^2}{1\over (\ell+1)^2} \; .\ee
This is also, famously, the value obtained in the Bohr model. 
As in \cite{FH1}, before we apply Bohr's correspondence principle, we check that in the large $\ell$ limit, the trial wave function -- like the exact result -- corresponds to strictly circular orbits, as in the Bohr model.  To that end, we compute the uncertainty in $r^2$, measured in units of mean square radius. We obtain 
\be \label{uncertainty} {[\langle r^4 \rangle _{\alpha \betaa \ell} -(\langle r^2 \rangle_{\alpha \betaa  \ell})^2]^{1\over 2}\over \langle r^2 \rangle_{\alpha \betaa \ell}}= 
\left ( {
{\Gamma \left ({2\ell +7\over \betaa} \right )\Gamma \left ({2\ell +3\over \betaa} \right )
\over {\Gamma \left ( {2\ell +5\over \betaa} \right )}^2}-1
} \right )^{1\over 2},\ee
a quantity which approaches 0 at large $\ell$ (see computation of the limit  in Appendix \ref{uncertaintycalc}), so the orbits are circular at large $\ell$.
In addition, since our trial system approximates a quantum system, it
has quantized energy and angular momenta. Therefore, our trial system satisfies the postulates of the (semi-classical) Bohr model: circularity of orbits and quantization of energy and angular momenta. 
According to Bohr’s correspondence principle, in the limit of large quantum number, quantum quantities approach classical ones \cite[p. 31]{Bo}, \cite[p. 117]{ER}. It now follows that in the limit $\ell \rightarrow \infty$,  our approximate quantum system approaches the classical model, as does the exact quantum system. Therefore, the ratio of the approximate value of the energy level given by equation (\ref{ExpectH}) to the classical (Bohr) value 
given by equation (\ref{exact}) approaches 1 in this limit. That is, 
\be \label{mainlimit} \lim_{\ell \rightarrow \infty}{\langle H \rangle _{min}^{\ell, b}\over   E_{0, \ell}}=\lim_{\ell \rightarrow \infty} 
{4\over \betaa }{(\ell+1)^2\over (2\ell + \betaa +1)}\; {\left (\Gamma({2\ell +2\over \betaa})\right )^2\over \Gamma ({2\ell +3\over \betaa})\Gamma ({2\ell +\betaa +1\over \betaa})}=1.
\ee
This limit leads to infinite product formulas for combinations of gamma functions of the form

\[ {\Gamma(x)\Gamma(y)\over \Gamma \left ( {x+y\over 2} \right )^2}\]
for certain values of $x$ and $y$ which we now discuss.

Let $\betaa$ be a positive even integer, and let us consider the subsequence of integers $\ell$ which are equivalent 
modulo $\betaa/2$, the reciprocal of the coefficient of $\ell$ in the arguments of $\Gamma$ in equation (\ref{mainlimit}); the limit of the expression in equation (\ref{mainlimit}) when the integers in the subsequence go to infinity is unity. 
Since we have $\ell \equiv a \mbox{ mod }{\betaa /2}$ for some $a\in \{0,1,2, \ldots , {b\over 2}-1\}$, we also have $\ell = a+{k\betaa\over 2}$ for some non-negative integer $k$.  Therefore,
\small
$$\Gamma \left ({2\ell +2\over \betaa} \right ) = \Gamma \left ({2a+2\over \betaa}+k \right ) ,
$$
\normalsize
and similarly for the other values of $\Gamma (x)$  that appear in equation (\ref{mainlimit}). 
We can now use the recursion relation $\Gamma (z+k)=(z+k-1)(z+k-2)\cdots (z+1)z\Gamma (z)$ and some algebraic manipulations to rewrite our limit as an infinite product formula for a combination of $\Gamma$ functions:
\be \label{infiniteproduct} {\Gamma \left ({2a+1+2\betaa \over \betaa} \right )\Gamma \left ({2a+3\over \betaa} \right )\over {\Gamma \left ( {2a+2+\betaa \over \betaa} \right )}^2}
= \prod_{k=1}^{\infty} {(k\betaa + 2a+2)^2\over (k\betaa +2a-\betaa +3)(k\betaa +2a+1+\betaa)} \; ,
\ee
or equivalently,
\be \label{infiniteproduct2} {\Gamma \left ({2a+1 \over \betaa} \right )\Gamma \left ({2a+3\over \betaa} \right )\over {\Gamma \left ( {2a+2 \over \betaa} \right )}^2}
= {(2a+2)^2\over (2a+1)(2a+1+\betaa )}\prod_{k=1}^{\infty} {(k\betaa + 2a+2)^2\over (k\betaa +2a-\betaa +3)(k\betaa +2a+1+\betaa)} \; .
\ee

When $\betaa =2$ and $a=0$, we have the case of \cite{FH1} and the above expression is the Wallis formula for $\pi$, equation (\ref{Wallis}).

\subsection*{Product formulas from the $N$-dimensional hydrogen atom}

The analogous computation in arbitrary dimensions leads to additional product formulas. The radial equation for the hydrogen atom in $N$ dimensions is \cite{Nieto}
\[ H_NR=\left [ -{\hbar ^2\over 2m}\left ( {d^2\over dr^2}+{N-1\over r}{d\over dr}-{\ell(\ell +N-2)\over r^2}\right ) -{e^2\over r} \right ] R(r)=ER(r),
\]
where $\hbar ^2 \ell (\ell+N-2)$, $\ell = 0,1,2, \ldots$  is the spectrum of the square of the angular momentum operator in $N$ dimensions \cite{Louck, FH2}. 
The same trial wave function as in three dimensions, equation (\ref{trial}), with the $Y^m_\ell (\theta, \phi)$ replaced by its $N$-dimensional analog \cite{Nieto, Louck} gives
\be \label{HNmin} \langle H_N \rangle ^{\ell, b}_{min}=  -{me^4\over 2\hbar ^2}{4\over b(2\ell +b+N - 2)} \left [ {\Gamma({2\ell +N-1\over b})^2\over \Gamma ({2\ell +N\over b})\Gamma({2\ell+b+N-2\over b})}\right ].\ee
The exact result  in $N$ dimensions is \cite{Nieto} 
 $$E^N_{n_r,\ell}=-{me^4\over 2\hbar ^2}{1\over (n_r+\ell+{N-1\over 2})^2}\; .
$$
In the limit $\ell \rightarrow \infty$ with $n_r=0$ we have

\be \label{mainlimit-N} \lim_{\ell \rightarrow \infty}{\langle H_N \rangle _{min}^{\ell, b}\over   E_{0, \ell}}=\lim_{\ell \rightarrow \infty} 
{4\over \betaa }{(\ell+{N-1\over 2})^2\over (2\ell + \betaa +N-2)}\; {\left (\Gamma({2\ell +N-1\over \betaa})\right )^2\over \Gamma ({2\ell +N\over \betaa})\Gamma ({2\ell +\betaa +N-2\over \betaa})}=1.
\ee
As before, we let $\betaa$ be a positive, even integer and  $\ell \equiv a \mbox{ mod }{\betaa /2}$. We then obtain the infinite product formula
\be \label{infiniteproduct2N} {\Gamma \left ({2a+N-2 \over \betaa} \right )\Gamma \left ({2a+N\over \betaa} \right )\over {\Gamma \left ( {2a+N-1 \over \betaa} \right )}^2}
= {(2a+N-1)^2\over (2a+N-2)(2a+N-2+\betaa )}\prod_{k=1}^{\infty} {(k\betaa + 2a+N-1)^2\over (k\betaa +2a-\betaa +N)(k\betaa +2a+\betaa+N-2)} \; ,
\ee
for $N\geq 3$. 

\vskip .5cm

\subsection*{Euler's reflection formula from product formulas} 
In our infinite product formula equation (\ref{infiniteproduct2}), let $a= {\betaa \over 2}-1$, with $\betaa$ a positive even integer as before. After using the identity $\Gamma (1+ {1\over \betaa})={1\over \betaa}\Gamma ({1\over \betaa})$, we have 
\be \label{reflectionproduct} \Gamma \left (1-{1\over \betaa}\right )\Gamma \left ({1\over \betaa }\right )={\betaa ^3\over (\betaa -1)(2\betaa -1)}
\prod _{k=1}^\infty  {(k\betaa +\betaa )^2\over (k\betaa +1)(k\betaa +2\betaa -1)} .\ee
Let $z={1\over \betaa}$, take the reciprocal of the above equation, and rewrite it as a power series in $z-1$. This equation then becomes
\[ {1\over \Gamma (1-z) \Gamma(z)}= -(z-1) \prod _{k=1}^\infty  \left ( 1-{(z-1)^2\over k^2}\right ).\]
The right hand side turns out to be Euler's formula for ${-\sin \pi (z-1) \over \pi}={\sin \pi z \over \pi}$, leading us to the formula
\be \label{Euler}\Gamma (1-z) \Gamma(z)={\pi \over \sin \pi z}.\ee
This formula is known as Euler's reflection formula and holds for any complex number $z$ that is not an integer. 

\subsection*{A few notable examples}
\begin{itemize}
\item The case $\betaa =4$ gives two formulas that were found by Catalan in 1873 \cite{Ca}: 
\begin{enumerate}
\item $\betaa=4$ and $a=1$ (i.e. odd $\ell$): equations (\ref{reflectionproduct}) and (\ref{Euler}) combine to give us the infinite product formula
\be {\pi \over 2\sqrt{2}}={4 \cdot 4\over 3\cdot 7}\prod _{k=1}^\infty  {(4k+4)^2\over (4k+1)(4k+7)} = \left ( {4 \cdot 4\over 3\cdot 7}\right ) {8\cdot 8\over 5\cdot 11}\; {12\cdot 12\over 9\cdot 15}\; {16\cdot 16\over 13\cdot 19} \cdots .\ee

\item $\betaa =4$ and $a=0$ (i.e. even $\ell$): setting $z={1\over \betaa}={1\over 4}$ in equation (\ref{Euler}) gives 
$$\Gamma \left ( {3\over 4} \right ) \Gamma \left ({1\over 4}\right ) = {\pi \over \sin{\pi/4}}=\pi \sqrt{2}.$$ Combining this with the identity 
$\Gamma ({1\over 2})=\sqrt{\pi}$ and equation (\ref{infiniteproduct2}) with $\betaa =4$ and $a=0$, we obtain the following infinite product formula for $\sqrt{2}$: 
\be 
\sqrt{2}={2\cdot 2\over 1\cdot 5} \prod _{k=1}^\infty  {(4k+2)^2\over (4k-1)(4k+5)}= \left ( {2\cdot 2\over 1\cdot 5}\right ) {6\cdot 6\over 3\cdot 9}\; {10\cdot 10 \over 7\cdot 13}\; {14\cdot 14\over 11\cdot 17}\cdots .\ee
\end{enumerate}

\item The case $\betaa =6$, $a=0$ gives

\[ \prod _{k=1}^\infty  {(6k+2)^2\over (6k-3)(6k+7)} = {7\sqrt{\pi}\over 4}{\Gamma \left ({1/6}\right )\over \Gamma \left ({1/3}\right )^2}. \]
The particular combination of $\Gamma$ functions on the right hand side appears in various places including Table 3 of \cite{BZ}; the right hand side also equals ${7\sqrt{3}\over 4 \cdot 2^{1/3}}$.

\item The case $\betaa =6$, $a=2$ gives another formula for $\pi$; equations (\ref{reflectionproduct}) and (\ref{Euler}) combine to give 
\be {\pi \over 3} = {6 \cdot 6\over 5\cdot 11}\prod _{k=1}^\infty  {(6k+6)^2\over (6k+1)(6k+11)} = \left ({6 \cdot 6\over 5\cdot 11}\right ) {12\cdot 12\over 7\cdot 17}\; {18\cdot 18\over 13\cdot 23}\; {24\cdot 24\over 19\cdot 29} \cdots .
\ee
\item The case $\betaa =8$, $a=3$ gives 
\be \label{eights} {\pi \over 4\sqrt{2-\sqrt{2}}} ={8 \cdot 8\over 7\cdot 15} \prod _{k=1}^\infty  {(8k+8)^2\over (8k+1)(8k+15)}=\left ( {8 \cdot 8\over 7\cdot 15}\right ){16\cdot 16 \over 9\cdot 23}\; {24\cdot 24 \over 17\cdot 31}\; {32\cdot 32\over 25\cdot 39}\cdots \; ,
\ee
where we used the identity $\sin \left ({\pi/8}\right )={\sqrt{2-\sqrt{2}}\over 2}$. Equation (\ref{eights}) coincides with formula (8) in \cite{SY}, where additional Wallis-like formulas are derived. 

\item The case $\betaa = 2^n$ and $a=2^{n-1}-1$ for an integer $n\geq 3$. Using the formula
$$\sin \left ( {\pi\over 2^n}\right )={1\over 2}\sqrt{2-\underbrace{\sqrt{2+\sqrt{2+\sqrt{2+ \cdots \sqrt{2}}}}}_\text{n-2 times}}\; , \hskip 1cm n\geq 3 , $$
we obtain
\be {2\pi \over \sqrt{2-\underbrace{\sqrt{2+\sqrt{2+\sqrt{2+ \cdots \sqrt{2}}}}}_\text{n-2 times}}} = {2^{3n}\over (2^n-1)(2^{n+1}-1)}
\prod _{k=1}^\infty  {(2^nk+2^n)^2\over (2^nk+1)(2^nk+2^{n+1}-1)}\; .
\ee
\item The case $N={b\over 2}-2a+1$ in equation (\ref{infiniteproduct2N}) leads to an interesting variation on the reflection formula. We get 
\[ {1\over \pi} \, \Gamma \left ({1\over 2}+{1\over \betaa}\right )\Gamma \left ({1\over 2}-{1\over \betaa }\right )={({\betaa \over 2}) ^2\over ({\betaa \over 2} -1)({3\betaa \over 2} -1)}
\prod _{k=1}^\infty  {(k\betaa +{\betaa \over 2})^2\over (k\betaa +1-{\betaa \over 2})(k\betaa +{3\betaa \over 2}-1)} .\]
Let $z={1\over 2}-{1\over \betaa }$, take the reciprocal of the above equation, and rewrite it in terms of $z+{1\over 2}$ to get
\[ {1\over \Gamma (1-z) \Gamma(z)}={ \left [4\left (z+{1\over 2}\right )^2 -1\right ] \over \pi} \prod _{k=1}^\infty  \left ( 1-{(z+{1\over 2})^2\over (k+{1\over 2})^2}\right ).\]

\end{itemize}

\subsection*{Relation to ideas of Brouncker and Ramanujan}

We can also relate the case of $b=2$ to a combination of a formula of William Brouncker and a formula of Srinivasa Ramanujan. See \cite{Kh, Du} for detailed historical accounts of the following. In 1655, as a result of correspondence from Wallis, Brouncker came up with a functional formula and an infinite continued fraction that is intimately related to Wallis's formula for $\pi$:
\begin{align}
 f(s-1)f(s+1)&={ s}^2 \; , \\  f(s)&=s+\frac{1^2}{2 s + \frac{3^2}{2{s} + \frac{5^2}{2{ s} +_{\ddots}}}}
\end{align}
Brouckner mentioned that $$f(1)={4\over \pi} \; .$$ 
Combining ideas of Brouncker and Ramanujan (Theorems 5 and 6 in \cite{Kh}), we have that  for every $s>0$, 
\begin{equation}\label{raman} f(s)=4 \left [{\Gamma \left ( {3+s\over 4}\right ) \over \Gamma \left ( {1+s\over 4}\right ) } \right ] ^2 = (s+1) \prod _{n=1}^\infty {(s+4n-3)(s+4n+1)\over (s+4n-1)^2}. 
\end{equation}
If we let $b=2$ and $a={s-1\over 4}$ in our product formula equation (\ref{infiniteproduct}), we obtain precisely the reciprocal of equation (\ref{raman}). The case $s=1$ ($a=0$) is then the Wallis formula for $\pi$. This demonstrates that -- as we shall see more generally in Appendix \ref{usingdefinitions} -- the validity of our product formula equation (\ref{infiniteproduct}) is not limited to the values of $b$ and $a$ for which we derived it. 

\vskip .5cm

\subsection*{Relation between definitions of $\Gamma (z)$ and the correspondence principle}

The gamma functions that appear in equations (\ref{ExpectH}), (\ref{min}), and (\ref{HNmin}) arise from the integrals used in the computation of the expectation value of the Hamiltonian and the integral formula definition for the gamma function,
\be \label{gammaintegraldefinition} \Gamma(z)=\int _0^\infty e^{-t} \, t^{z-1} dt  ,\hskip 1cm \mbox{Re}(z)>0 \; .\ee
Then, the correspondence principle gives us the limit in equation (\ref{mainlimit}) or (\ref{mainlimit-N}), which leads to our infinite product formulas and in turn to Euler's reflection formula. Can the same infinite product formulas be obtained in any other way? The answer is ``yes," but
Euler's limit-based definition of the gamma function,
\be \label{gammalimitdefinition} \Gamma (z)=\lim _{m\rightarrow \infty} {m^z m! \over z(z+1)(z+2)\cdots (z+m)}\, , \hskip 1cm  z\in \mathbf C , \; z\neq 0,-1,-2, \ldots \; ,\ee 
is required. The details are carried out in Appendix \ref{usingdefinitions}. 

The mathematical equivalence of Euler's limit-based definition of the gamma function, equation (\ref{gammalimitdefinition}), and the integral definition, equation (\ref{gammaintegraldefinition}), requires proof (see \cite{Ar}), and is necessary for the mathematical derivation of the infinite product formulas as shown in Appendix \ref{usingdefinitions}. 
In our derivation, 
we do not need the limit-based definition. Bohr's correspondence principle provides the crucial link between the integral definition and the infinite product formulas we obtained. 

\vskip 1cm
\large
\noindent {\bf Acknowledgements}
\normalsize
\vskip .2cm

The authors would like to thank Thomas J. McElmurry and Carl Hagen for helpful discussions, to  Maxim Derevyagin for bringing \cite{Kh} to our attention, and to Jean-Paul Allouche for bringing \cite{WW} to our attention. We would also like to thank the editor and the anonymous referee for helpful comments.
The authors gratefully acknowledge the National Science Foundation (NSF MCTP-1143716) and Smith College for their support of the Center for Women in Mathematics at Smith College, where some of this work was completed. 
Some of this work was completed while the first author was at Haverford College. 

\vskip 1cm
\large
\noindent {\bf Data Availability Statement}
\normalsize
\vskip .2cm

Data sharing is not applicable to this article as no new data were created or analyzed in this study.

\vskip 1cm

\appendix 

\numberwithin{equation}{section}

\section{Large $\ell$ limit of uncertainty in $r^2$}\label{uncertaintycalc}
Here we show that the expression in equation (\ref{uncertainty}) approaches 0 at large $\ell$. 
From equation (\ref{gammalimitdefinition}), together with
the recursion relation $\Gamma(z+1)=z\Gamma(z)$ and the value $\Gamma (m+1)=m!$, we have
\[ 
\lim _{m\rightarrow \infty}{m^z\Gamma(m+1)\over \Gamma(z+m+1)}=1 ,
\]
from which it also follows (by taking an appropriate product of the above limit with $z=z_1, z_2, w_1, w_2$) that for $z_1+z_2=w_1+w_2$, \be
\label{AlsoUsefulLimit} 
\lim _{m\rightarrow \infty} {\Gamma(z_1+m)\Gamma (z_2+m)\over \Gamma(w_1+m)\Gamma (w_2+m)}=1 .
\ee

Now let $k=\lfloor {2\ell +3\over \betaa}\rfloor$, $r={2\ell +3\over \betaa}-k$, $s={2\ell +5\over \betaa}-k$, and $t={2\ell +7\over \betaa}-k$ (where $\lfloor x \rfloor$ is the greatest integer less than or equal to $x$). We have $t+r=2s$. Using equation (\ref{AlsoUsefulLimit}) with $z_1=t$, $z_2=r$, $w_1=w_2=s$, we have 
\[ {\Gamma \left ({2\ell +7\over \betaa} \right )\Gamma \left ({2\ell +3\over \betaa} \right )
\over {\Gamma \left ( {2\ell +5\over \betaa} \right )}^2}
={\Gamma  (t+k)\Gamma (r+k)\over {\Gamma (s+k)}^2}
\xrightarrow[{k\rightarrow \infty}]{} 1, \]
from which the result follows.

\section{Mathematical derivation of product formulas}\label{usingdefinitions}

First, we note that the finite version of the product on the right hand side of equation (\ref{infiniteproduct}) can be rewritten as
\[ \prod_{k=1}^{\ell} {(k\betaa + 2a+2)^2\over (k\betaa +2a-\betaa +3)(k\betaa +2a+\betaa+1)}= 
{\left ( 1+{2a+2\over \betaa}\right )^2_\ell \over \left ( {2a+3\over \betaa}\right ) _\ell \left ( 2+{2a+1\over \betaa}\right )_\ell} \; ,
\]
where $(y)_\ell=y(y+1)(y+2)\cdots (y+\ell-1)=\Gamma (y+\ell)/\Gamma(y)$ is the Pochhammer symbol. In turn, the expression on the right hand side becomes
\be
\label{bigproduct}
{\Gamma {\left ( 1+{2a+2\over \betaa}+\ell \right )}^2\over \Gamma {\left ( 1+{2a+2\over \betaa}\right )^2}}
{\Gamma \left ( {2a+3\over \betaa}\right ) \over \Gamma \left ( {2a+3\over \betaa}+\ell \right ) }
{\Gamma \left ( 2+{2a+1\over \betaa}\right )\over \Gamma \left ( 2+{2a+1\over \betaa}+\ell\right )}.
\ee

Now let $z_1= z_2= 1+{2a+2\over \betaa}$, $w_1={2a+3\over \betaa}$, $w_2= 2+{2a+1\over \betaa}$ in equation (\ref{AlsoUsefulLimit}), which follows from Euler's limit-based definition of the gamma function, equation (\ref{gammalimitdefinition}). We have
\be \label{limitratio} \lim _{\ell\rightarrow \infty} 
{\Gamma {\left ( 1+{2a+2\over \betaa}+\ell \right )}^2\over 
\Gamma \left ( {2a+3\over \betaa}+\ell \right )
 \Gamma \left ( 2+{2a+1\over \betaa}+\ell\right )}=1.
\ee
Putting equation (\ref{bigproduct}) together with the limit in equation (\ref{limitratio}), we obtain the product formula of equation (\ref{infiniteproduct}).

 Similarly,  we can obtain the product formula of equation (\ref{infiniteproduct2N}) by following the same procedure, with $z_1= z_2= 1+{2a+N-1\over \betaa}$, $w_1={2a+N\over \betaa}$, $w_2= 2+{2a+N-2\over \betaa}$. 
 
 Note that it follows that the product formulas (\ref{infiniteproduct}) and (\ref{infiniteproduct2N})  which we derived for positive even integers $\betaa$ and  for $a\in \{0,1,\ldots , {b\over 2}-1\}$, actually hold far more generally.
 
An even more general version of our product formulas is derived in \cite[Section 12.13]{WW} and can be stated as follows. If $d$ is a positive integer and $a_1 + a_2 + \cdots + a_d = b_1 + b_2 + \cdots + b_d$, where the $a_j$ and $b_j$ are complex numbers and no $b_j$ is zero or a negative integer, then
$$
\frac{\Gamma(b_1) \cdots \Gamma(b_d)}{\Gamma(a_1) \cdots \Gamma(a_d)}
=\prod_{n \geq 0}\frac{(n+a_1) \cdots (n+a_d)}{(n+b_1) \cdots (n+b_d)}.
$$
The derivation uses the Weierstrass definition of the gamma function:
\[ \frac{1}{\Gamma (z)}=ze^{\gamma z} \prod _{n=1}^\infty \left \{ \left (1+\frac{z}{n}\right ) e^{-z/n}\right \},
\]
where $\gamma$ is the Mascheroni constant. We are grateful to Jean-Paul Allouche for bringing this fact to our attention.

\end{document}